\documentclass[%
 reprint,
 amsmath,amssymb,
 aps,
 prl,
]{revtex4-2}

\usepackage{graphicx}
\usepackage{dcolumn}
\usepackage{bm}

\begin{document}

\preprint{APS/123-QED}

\title{Laser Measurement of Anomalous Electron Diffusion in a Crossed-Field Plasma}

\author{Parker J. Roberts}
\email{pjrob@umich.edu}

\author{Benjamin A. Jorns}%
 \email{bjorns@umich.edu}
\affiliation{%
University of Michigan, Ann Arbor, MI 48109, USA
}

\date{\today}

\begin{abstract}
Non-classical electron diffusion in crossed-field, low-temperature plasmas is measured experimentally.
Laser-induced fluorescence and Thomson scattering are used to determine the inverse Hall parameter, a metric for cross-field transport, in a Hall ion source. 
The measured diffusion is found to depart from fluid model results at the peak electric field, remaining constant instead of exhibiting a sharp transport barrier. 
The implications of this result are discussed in terms of the current understanding of non-classical diffusion in low-temperature plasmas.
\end{abstract}

\maketitle

Electron diffusion across confining magnetic fields is exhibited by nearly all types of magnetized plasma \cite{graham2022direct, taylor1961diffusion,curreli2014cross,carreras2005plasma,bell2019transport}.
Classically, this cross-field migration results from collisions with heavier species.  
In practice, however, such collisions are often insufficient to explain the measured rates of cross-field electron flux \cite{corkum_anomalous_1973,urrutia_anomalous_1986,shukla_cross-field_1987,podesta_cross-field_2008,hara_cross-field_2020}.  
This ``anomalous” transport can be orders of magnitude higher than the classically-predicted rate.    
Enhanced diffusion poses a particular challenge for the low-temperature, crossed-field plasma devices - such as Hall and Penning discharges - that are widely employed for commercial and research applications \cite{boeuf2017tutorial,phillips1899action}.  
It has been shown in these devices that the high electron transport adversely impacts the efficiency and ion beam quality, while also curtailing the development of fully predictive models \cite{goebel2023fundamentals,anders2012drifting}.  
These  practical considerations have motivated an extensive body of investigations into electron transport in such systems \cite{meezan_anomalous_2001, meezan_kinetic_2002,carlsson2018particle,anders2012drifting,brown_growth_2023,hecimovic2016anomalous}.
Despite these efforts, however, the process remains poorly understood. 

One of the major impediments for advancing  understanding of this problem stems from the sparsity of experimental data on cross-field electron diffusion.
While global consequences of the transport can be measured, such as the fractional current carried by electrons in the far field \cite{dannenmayer2013electron}, locally resolving electron diffusion has been largely intractable to date.  
Standard physical probing methods have historically proven to be ineffective in the near field, as they are too perturbative or cannot resolve the directionality of electron drift \cite{jorns2015plasma,grimaud2016perturbations}; moreover, non-intrusive experimental techniques that have been used to date to estimate the diffusion require strong and potentially unphysical assumptions \cite{thomas_nonintrusive_2006,dale_non-invasive_2019}. 
In light of these challenges, it has become common practice to use calibrated models to infer the diffusion locally.
This can be done, for example, by treating the electron diffusion with an effective transport coefficient, e.g. an anomalous collision frequency \cite{fife1998hybrid,davidson1977cross}. 
This parameter is then prescribed locally as a function of position and used to solve the governing equations in a multi-fluid or hybrid model for the discharge.  
The  values of the transport coefficient are adjusted iteratively until predictions from the model agree with more tractable, non-intrusive experimental measurements, such as the ion velocity. 
This approach has been applied extensively to magnetized plasmas \cite{kruer1972anomalous,choueiri1999anomalous,mikellides2016hall2de,tummel2020kinetic,marks_evaluation_2023,hecimovic2016anomalous}. With that said, this indirect approach has several limitations, including questions about the fidelity and uniqueness of the resulting inferred transport profile \cite{mikellides2019challenges,marks2023challenges}. In light of these challenges, there is a pressing need to directly and non-invasively infer the electron diffusion in these types of poorly-understood, crossed field plasmas.

In this letter, we present direct measurements of the anomalous cross-field electron diffusion, as quantified by an effective transport coefficient, based on a combination of two non-invasive, laser diagnostics: laser-induced fluorescence and incoherent Thomson scattering.  
We perform this experiment in one of the most common low-temperature, crossed-field plasma devices: a Hall effect accelerator. 
We then compare the measurement to a previously-reported indirect estimate for anomalous transport from a calibrated model.

\begin{figure}
    \centering
    \includegraphics[width=1\linewidth]{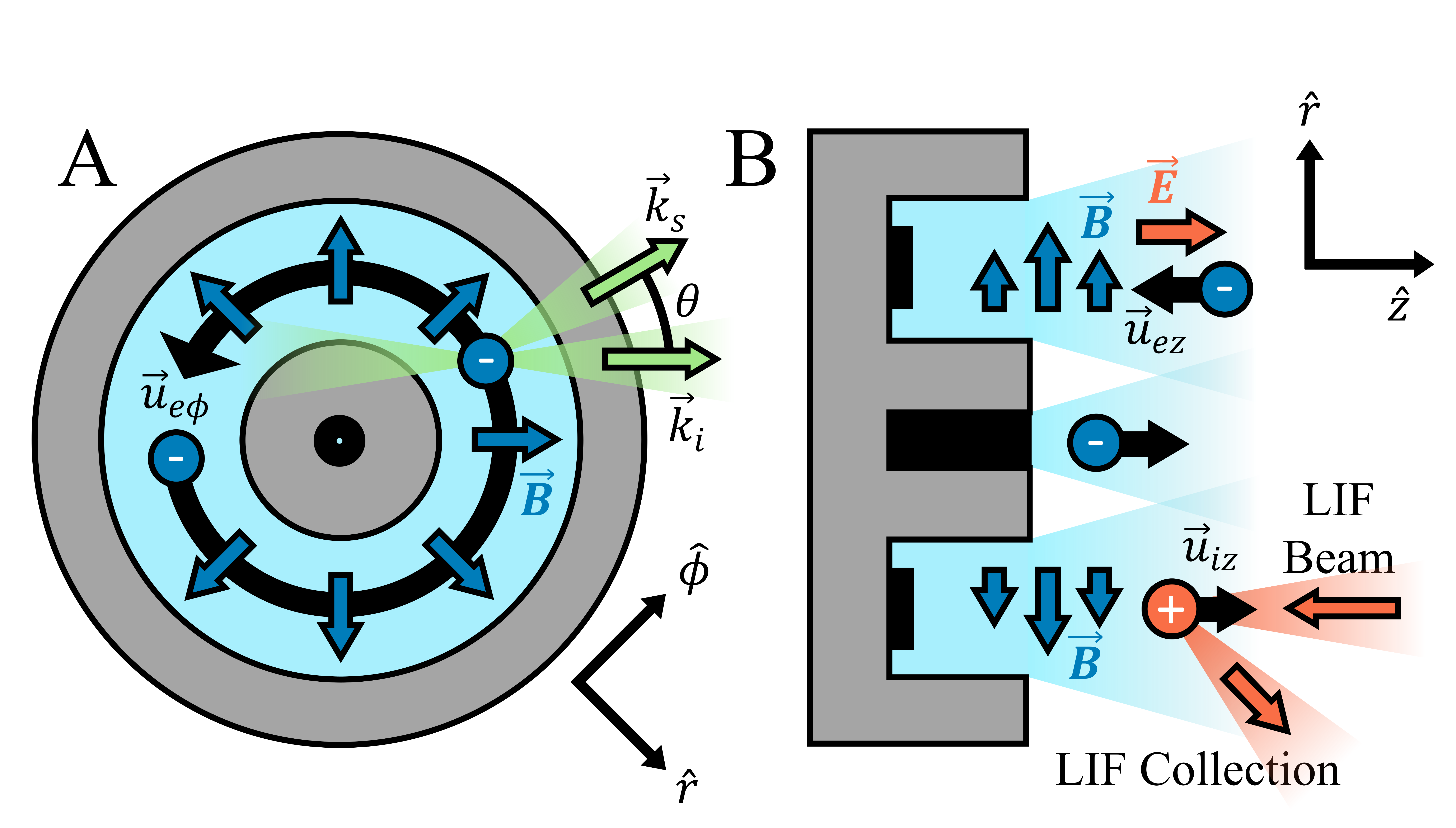}
    \caption{Illustration of a Hall accelerator with the employed laser scattering geometry.
    A) Front view with azimuthal Thomson scattering wavevectors.
    B) Side view with axial laser-induced fluorescence injection scheme.}
    \label{fig:HET_diagram}
    \vspace{-15pt}
\end{figure}

To motivate our approach, we show in Fig. \ref{fig:HET_diagram} a schematic of a canonical, axisymmetric Hall accelerator. This device features an annular plasma channel subject to a radial magnetic field $\vec{B} = B_r \hat{r}$ crossed with an axial electric field $\vec{E} = E_z \hat{z}$.  
The magnetic field strength is tailored so that the ions are effectively unmagnetized and accelerated by the electric field, while the light electrons are magnetized.
These electrons exhibit drifts that can be described by a generalized Ohm's law (neglecting electron inertia),
\begin{align}
    u_{e\phi}  = \frac{1}{\Omega^{-2} + 1} u_{\text{drift}}, & &
    u_{ez} = - \frac{\Omega^{-1}}{\Omega^{-2} + 1} u_{\text{drift}},  \label{eq:axial_velocity}
\end{align}
where  $ u_{\text{drift}} = E_z/{B_r} + [en_e B_r]^{-1} {\partial_z \left( n_e k_B T_e \right)}$ denotes the ideal, collisionless azimuthal drift arising from the electric field and diamagnetic effects, $e$ is the magnitude of the electron charge, $n_e$ is the electron density,  $T_e$ is the electron temperature, and $\Omega$ is the Hall parameter representing the ratio of cyclotron frequency,  $\omega_{ce} = eB_r/m_e$, to effective electron collision frequency (both classical and non-classical) \cite{hagelaar2011plasma}.
Physically, the relations in Eq.~\ref{eq:axial_velocity} illustrate how  the inverse Hall parameter, $\Omega^{-1}$, can be interpreted as an effective transport coefficient.  
With increasing values of $\Omega^{-1}$,  the azimuthal Hall drift, from which the accelerator derives its name, is reduced, while the cross-field drift is enhanced.
We subsequently focus on experimentally characterizing this transport coefficient.

To this end, we consider three ways to relate the inverse Hall parameter to measurements of the background plasma properties:
\begin{equation}
\begin{aligned}
     \Omega_A^{-1}  = \frac{ - 2u_{ez}/u_{\text{drift}}}{\left(1 + \sqrt{1 + 4 u_{ez}/u_{\text{drift}}} \right) }, \\
     \Omega_B^{-1}  = - \frac{u_{ez}}{u_{e \phi}}, \quad
     \Omega_C^{-1}  = \sqrt{\frac{u_{\text{drift}}}{u_{e\phi}} - 1},
     \label{eq:Hallparams}
\end{aligned}
\end{equation}
Method ``C" is not feasible experimentally, because the small value of $\Omega^{-1} \ll 1$ in strongly magnetized plasmas makes $u_{\text{drift}}$ and $u_{e\phi}$ nearly indistinguishable; we thus only consider methods ``A" and ``B" in this work.
Evaluating $\Omega^{-1}$ from these two formulae is possible given the electron density, temperature, and velocity along the azimuthal and axial directions, in addition to the electric field.

We characterized in this work these plasma properties along the channel centerline of the H9, a 9-kW class, magnetically shielded Hall effect accelerator \cite{hofer2017h9}, operating at a discharge voltage of 300 V and 15 A on krypton.  This device has been the subject of extensive experimental study, including efforts to indirectly infer transport from calibrated simulations \cite{su_investigation_2022,su_trends_nodate}. We tested this device in the Alec D. Gallimore Large Vacuum Test Facility, a 6-meter-diameter and 9-meter-length vacuum chamber maintained at $5$ \textmu Torr during operation \cite{viges2019university}. 

\begin{figure}
    \centering
    \includegraphics[width=1\linewidth]{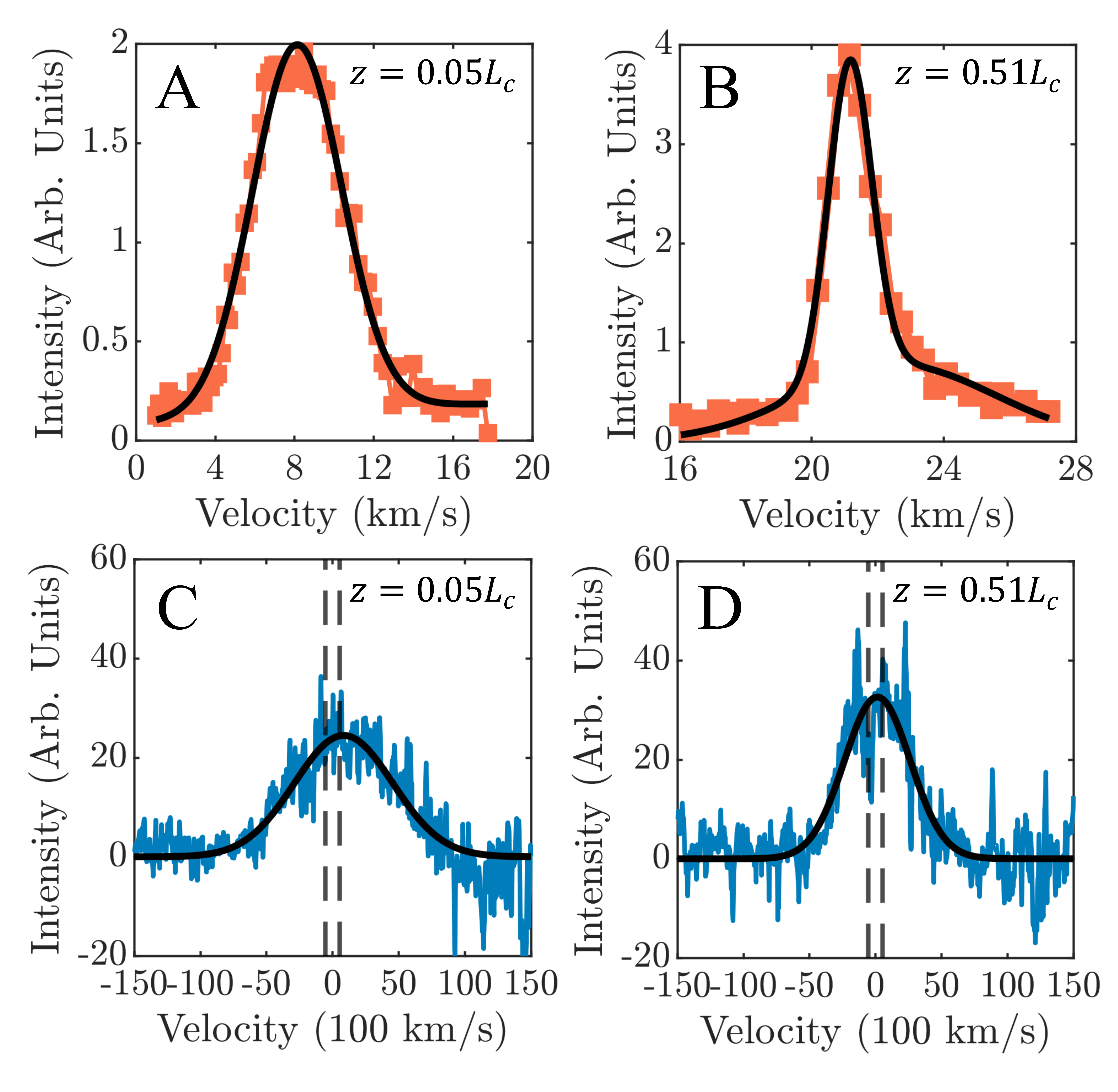}
    \caption{A-B) Axial ion velocity distributions from laser-induced fluorescence with model fit, where $L_c$ is the H9 channel length. C-D) Azimuthal electron velocity distributions from incoherent Thomson scattering with Maxwellian fit, showing notch filter stopband (dashed lines).}
    \label{fig:Injection}
      \vspace{-15pt}
\end{figure}

To infer the axial electric field, $E_z$, we examined the axial ion velocity distribution function (IVDF) along the channel centerline.   
Following the method of Perez-Luna et al. \cite{perez2009method}, the first three moments of this IVDF yielded estimates for the local electric field. 
We measured these distribution functions with the laser-induced fluorescence (LIF) system described in Ref. \cite{su_trends_nodate}, which is based on the principle of characterizing the Doppler-broadened lineshape of a laser-induced transition from a metastable state of singly-charged krypton ions.  
As discussed in Refs. \cite{su_trends_nodate} and \cite{dale2019investigation}, non-Doppler broadening effects can be neglected for this regime such that the LIF lineshapes approximate the IVDF along the laser wavevector. 
Figure \ref{fig:HET_diagram}B illustrates the injected beam and collection optic relative to the H9, which intersected to form an interrogation point with 1-mm spatial resolution.
We moved this point relative to the Hall accelerator by translating the device on a motion stage. 
Figures \ref{fig:Injection}A-B show IVDFs inferred from LIF (with a velocity resolution of 200 m/s) at two locations along channel centerline.
Fits to a sum of two Gaussian distributions (solid lines) were used for analysis of the distributions.

We used incoherent Thomson scattering (ITS) to directly measure the electron properties $u_{e\phi}$, $n_e$, and $T_e$ along channel centerline. 
This diagnostic measures the electron velocity distribution function (EVDF), inferred from the plasma light scattering spectrum.
As shown in Fig. \ref{fig:HET_diagram}A, in our ITS setup, we focused a  Q-switched, Nd:YAG laser, with wavelength $\lambda_0 = 532$ nm and a pulse energy of 700 mJ, into the plasma with wavevector $\vec{k}_{i}$. 
An in-situ optic collected scattered radiation at wavevector $\vec{k}_{s}$, angled $\theta = 30^{\circ}$ from $\vec{k}_i$, providing 1-mm spatial resolution.
This optic routed light via an optical fiber to a detection bench, described in Ref. \cite{roberts2023cathode}, consisting of a three-stage volume-Bragg-grating-based stray light filter \cite{vincent_compact_2018}, a spectrometer, and an EMICCD camera.
We averaged 3000 spectra at each location, subtracting the background plasma emission.
In the incoherent scattering regime, which is satisfied for our plasma conditions and laser wavelength \cite{sheffield2010plasma}, the  power scattered at wavelength $\lambda = 2\pi/k_s$ is proportional to the number of electrons with velocity, $v$, projected along the scattering vector, $\Delta \vec{k} = \vec{k}_{s} - \vec{k}_{s}$.
This proportionality is governed by a Doppler shift: $v(\lambda) = c\left( \lambda_0/\lambda - 1 \right)/\left( 2 \sin \left[ \theta/2 \right] \right)$,
where $c$ is the speed of light and $\lambda_0$ is the laser wavelength, corresponding to a velocity resolution for our system of 29 km/s.
A motion stage translated the measurement region axially, with $\Delta \vec{k}$ aligned to measure azimuthal velocities.

Figures \ref{fig:Injection}C-D show azimuthal EVDFs at two locations, with the notch filter bounds included for reference.   
We extracted electron properties from these spectra by regressing a model for the convolution of a Maxwellian velocity distribution with the measured instrument broadening function, $I(\lambda)$ \cite{roberts2023cathode,vincent_compact_2018}:
\begin{equation}
    g(\lambda_k) = \sum\limits_{\ell=1}^{N} \frac{H n_e r_e^2}{ v_{Te} \sqrt{\pi}} e^{ \left( - \frac{ \left(v(\lambda_\ell) - u_e \right)^2}{v_{Te}^2} \right) } I\left( \lambda_k - \lambda_\ell \right),
    \label{eq:model}
\end{equation}
where $v_{Te} = \sqrt{2 k_B T_e/m_e}$ is the thermal velocity, $r_e$ denotes the classical electron radius, and $H$ is an intensity calibration factor we determined from rotational Raman scattering on nitrogen.  We show as bold lines in Fig. \ref{fig:Injection}C-D fits of Eq. \ref{eq:model} to the ITS spectral data outside of the notch filter bandwidth.

The prescription in Eq.~\ref{eq:axial_velocity} depends on the axial component of electron velocity, $u_{ez}$, while we measured only the azimuthal component.
We estimated $u_{ez}$ from the other plasma measurements by invoking local current continuity, i.e. $u_{ez} = u_{iz} - J_z/(e n_e)$, where $J_z$ is the total axial current density on centerline.
This prescription assumes symmetry about channel centerline and ignores radial velocity gradients.
The symmetry and collimation of ion velocities have been validated extensively in previous LIF studies of  Hall accelerators (c.f. Refs. \cite{hargus2001laser,duan2020investigation}).
In turn, extensive modeling and experiments suggest negligible centerline electron current density in the radial direction, which lies along field lines.\cite{mikellides2014magnetic,hofer2014magnetic,garrigues2003model,morozov2000fundamentals}.  
In order to estimate $J_z$, we used the conversion $\eta_b J_z = e n u_{iz}|_{end}$, where $\eta_b$ is the so-called ``beam utilization efficiency" of the device and end denotes the furthest downstream measurement; this efficiency was determined to be  $\eta = 0.83 \pm 0.04$ for this device and operating condition in previous work \cite{su2021performance}.

\begin{figure}
    \centering
    \includegraphics[width=\linewidth]{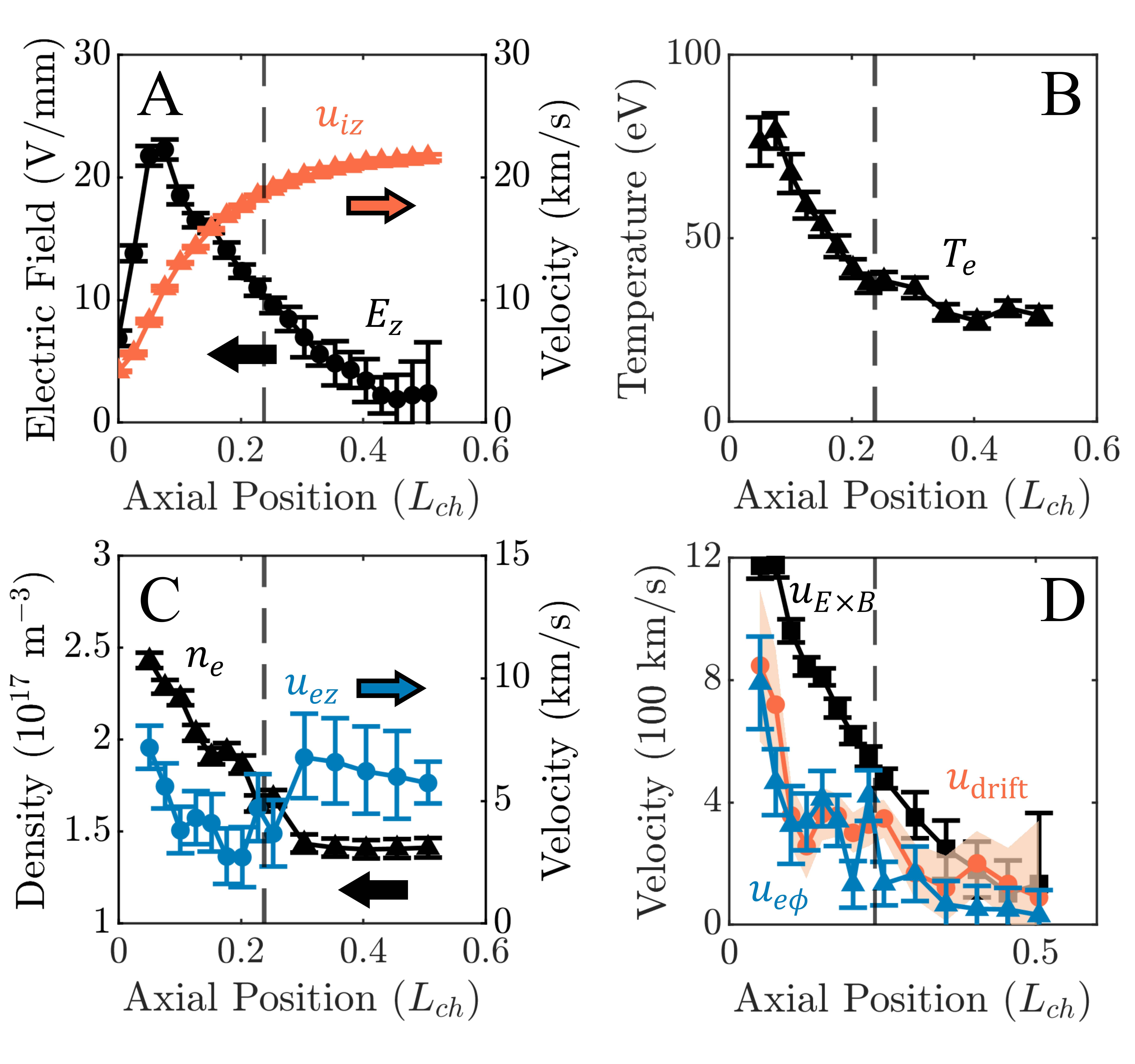}
    \caption{Channel centerline plasma properties, with 95\% credible intervals and magnetic field peak (dashed line). A) Axial ion velocity and electric field; B) Azimuthal electron temperature; C) Electron density and axial electron velocity magnitude; D) Azimuthal electron velocity compared to $u_{E\times B}$ and $u_{\text{drift}}$.}
    \label{fig:Results}
    \vspace{-15pt}
\end{figure}

Armed with the LIF and ITS techniques, we show in Fig.~\ref{fig:Results} the evolution of the time-averaged plasma properties along channel centerline, with distance normalized by the channel length $L_{ch}$.
We display 95\% credible intervals from the least-squares fit error.  
Fig.~\ref{fig:Results}A shows the ion velocity and electric field, demonstrating acceleration commensurate with the applied discharge voltage (300 V).
The maximum magnetic field (dashed line) serves as a barrier to electron motion, inducing a maximum in the electric field within 0.1 channel lengths to support the discharge current (c.f. Ref.~\cite{goebel2023fundamentals}).

We show in Fig.~\ref{fig:Results}B and C the azimuthal electron temperature, density, and axial velocity.
The large electric field Ohmically heats electrons, while ion acceleration causes a decrease in the plasma density; therefore both $T_e$ and $n_e$ decrease with $E_z$.
Meanwhile, the cross-field electron drift of 2.5-6 km/s contributes 20\% of the plasma current.
Notably, the peak temperatures we measure, $\sim 80$ eV, are 2-2.5 times larger than both channel wall probe measurements \cite{hofer2014magnetic} and predictions from models  \cite{su_investigation_2022,mikellides2016hall2de} of similarly configured ion sources.  
However, the magnitude of these measured temperatures is consistent with other Thomson scattering investigations of lower-power Hall accelerators \cite{vincent_incoherent_2020}.
This significant discrepancy with previous probe results may suggest that probe-based methods thermally perturb the plasma \cite{jorns2015plasma,grimaud2016perturbations}.
In turn, the contrast with modeling findings may indicate potential shortcomings in assumptions used to model electron energy transport \cite{marks2023challenges,marks_evaluation_2023}, which we revisit in the following.

We show in Fig. \ref{fig:Results}D the directly measured azimuthal electron drift, $u_{e \phi}$.
Overall, the magnitude of the drift is on the order of 100-800 km/s, approximately 10\% of the electron thermal speed and 10-100 times higher than the axial drift.  
For comparison, we also show the $E\times B$ drift and the ideal, collisionless drift $u_{\text{drift}}$, which accounts for the diamagnetic drift.
We estimate the latter by equating the axial temperature with the measured azimuthal temperature - an assumption we return to in the following.
With that said, Fig. \ref{fig:Results}D shows that the $E\times B$ drift is maximized with the peak electric field, while the total collisionless drift, $u_{\text{drift}}$, is reduced from this value due to the negative diamagnetic contribution.
The measured drift, $u_{e \phi}$, agrees with $u_{\text{drift}}$ to within uncertainty, consistent with an inverse Hall parameter much smaller than unity (Eq.~\ref{eq:axial_velocity}).
This agreement motivates a posteriori our decision to refrain from using method ``C'' for inferring the Hall parameter.

Having established the pertinent plasma measurements, we can now infer the inverse Hall parameter from both the diamagnetic drift and $E\times B$ drift (method ``A'') and directly from the axial and azimuthal drifts (method ``B''). 

\begin{figure}
    \centering
    \includegraphics[width=\linewidth]{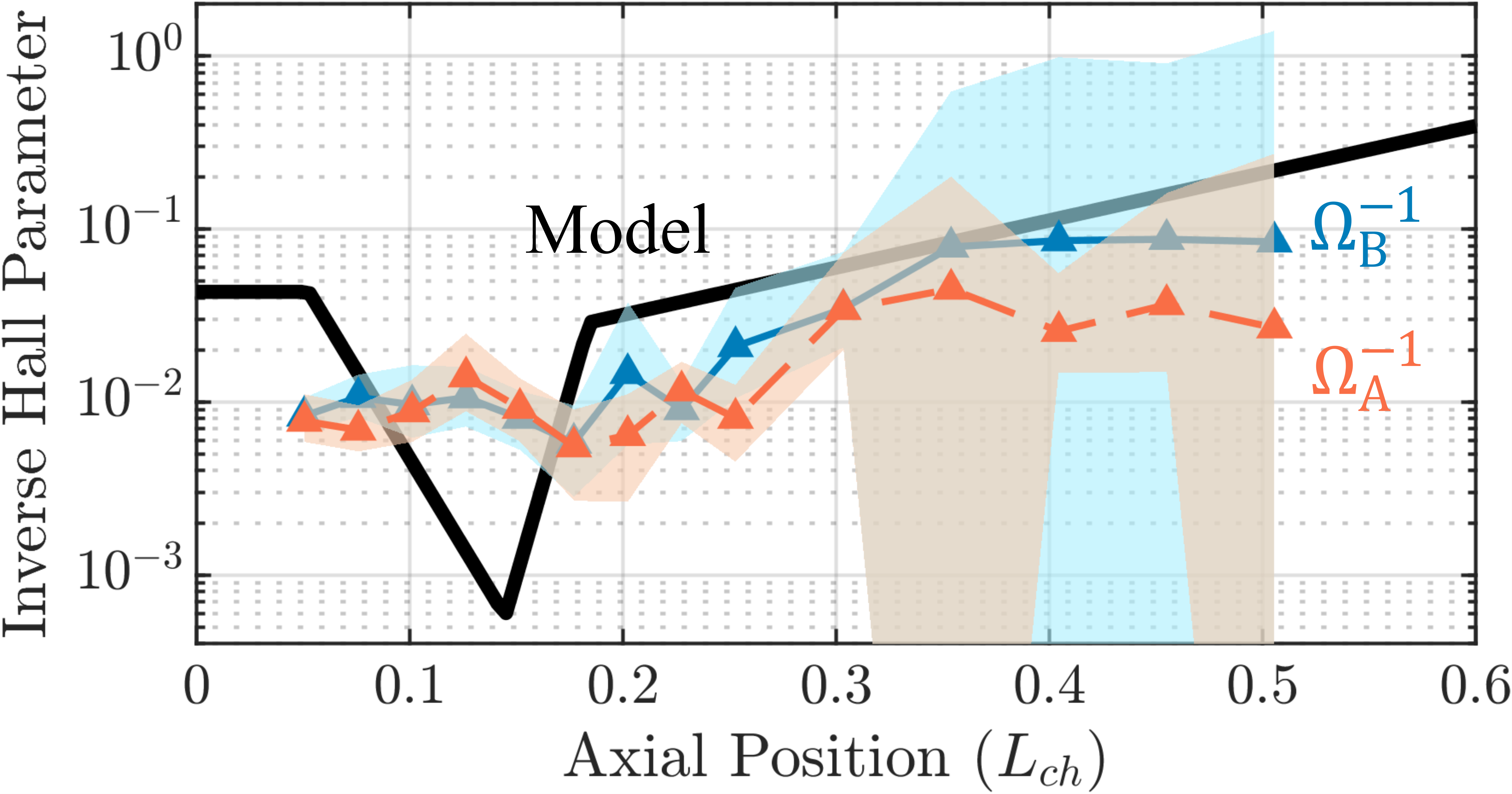}
    \caption{Inverse Hall parameter measurements with 95\% credible intervals from methods ``A" and ``B."  The solid line corresponds to the inverse Hall parameter calibrated for simulation results from Ref. \cite{su_investigation_2022}.}
    \label{fig:collision_frequency}
    \vspace{-15pt}
\end{figure}

We show the resulting median estimates with 95\% credible interval bounds propagated from the fit uncertainties in Fig. \ref{fig:collision_frequency}.
We note here several features from these two methods. 
First, the credible intervals overlap over the domain.
This lends confidence to both methods and helps address concerns about the use of azimuthal temperature in lieu of axial temperature in method ``A." 
Second, we note that at multiple locations, there is a lower bound of zero on the uncertainty estimates, which occurred when $u_{e\phi}$ and $u_{\text{drift}}$ were within uncertainty of zero (c.f. Fig. \ref{fig:Results}D).  
Third, we see that the median inverse Hall parameter estimates are nearly constant at $\Omega^{-1} \sim 0.01$ where $E_z$ is large ($z < 0.2 L_{ch}$), then increase to $\Omega^{-1} \sim 0.04-0.08$ at the downstream boundary.   
This relatively small upstream value is qualitatively consistent with  previous experimental and numerical studies of these devices \cite{meezan_anomalous_2001,dale_non-invasive_2019,mikellides2016hall2de,marks2023challenges}.  

The results in Fig. \ref{fig:collision_frequency} represent, to our knowledge, the most direct and non-intrusive measurements of cross-field transport in this class of device performed to date. 
As such, these results represent a ``ground truth" for evaluating the model-based methods commonly used to infer the inverse Hall parameter. 
To this end, we show for comparison in Fig. \ref{fig:collision_frequency} values of the inverse Hall parameter determined from a model for this device and operating condition reported in previous work \cite{su_investigation_2022,su_trends_nodate}. 
In that work, these values were iteratively tuned until outputs of a multi-fluid simulation matched LIF-based ion velocity data.   
As can be seen from that result, the magnitude of the simulated and experimental results qualitatively agree. 
Moreover, the two methods scale similarly in the region downstream of the peak magnetic field ($z/L_{ch} > 0.2$), which lends validation to the indirectly inferred profile.  

With that said, there is a notable discrepancy that arises from this comparison: close to the peak in electric field, the simulation-based result differs from the nearly constant experimental values by an order of magnitude, instead exhibiting a sharp minimum.
This minimum, which causes a localized barrier to cross-field transport, is necessary to capture the strong electric fields and rapid spatial acceleration of ions from experimental measurement \cite{marks2023challenges,lafleur2016theory,lafleur2016theory2,mikellides2016hall2de}.  
Our experiment suggests that in the real system, however, this transport barrier is not as stark as the simulation assumes, and is instead characterized by a wider and flatter transport curve. 

Our results, then, invite the question as to how the system can physically support the measured electric fields in the absence of the strong transport barrier inferred from simulation. 
An explanation stems from another departure our results make from the simulation, which we noted in the preceding: the disparity between measured and simulated electron temperatures.
As can be inferred from Eq. \ref{eq:axial_velocity}, in the limit of small inverse Hall parameter and for $u_{ez} < 0$, the electric field scales as $E_z = - \Omega B_r u_{ez} - \left( e n_e \right)^{-1} \partial_z(n_e k_B T_e)$. 
In other words, for a fixed crossed-field drift, the electric field is moderated by the effective cross-field impedance, which scales with $\Omega$. 
There are consequently two possible ways to result in an increased electric field strength for a given cross-field drift: a lower inverse Hall parameter, or a stronger pressure gradient.
Thus, even though the inverse Hall parameter is larger than the simulation-based result, the plasma can still maintain a strong electric field by virtue of the higher experimentally observed temperatures. 
With that said, the reason why models may underpredict the electron temperature by roughly a factor of two is an open question.
Recent work has suggested that this might be attributed to limitations in how models treat non-classical energy diffusion \cite{marks_evaluation_2023}. 
It is a common practice, for example, to represent the anomalous heat flux with a Fourier law based on the same value of the non-classical inverse Hall parameter inferred for the electron drift \cite{mikellides2012numerical,chapurin2022fluid}.  
The theoretical underpinnings of this approximation are not well-supported - a fact which is demonstrated by comparing our experimental temperature measurements to model results.

As a concluding remark, we comment here on limitations of our method.  
First, our measurements are time-averaged, whereas previous studies have suggested that $\Omega^{-1}$ could vary significantly at frequencies of 10-100 kHz \cite{dale2019investigation,charoy2021interaction}. 
Our conclusions about the large electron temperature and shallowness of the time-averaged transport barrier may not apply equally to all phases of these fluctuations.
However, models often assume a steady diffusion profile as well.
Second, both our experimental approach and most fluid models do not account for effects of electron inertia, though some work does suggest that these effects could play a non-negligible role in transport physics \cite{sahu2020full}.
Even with these caveats, however, these findings represent new and direct insight into time-averaged particle diffusion across magnetic fields. 
The departures from our current understanding of this transport therefore invite a reconsideration of the assumptions used to model momentum and energy transport in low-temperature plasmas.

\begin{acknowledgements}
This work was supported by a NASA Space Technology Graduate Research Opportunity (Grant 80NSSC20K1229), the Air Force Office of Scientific Research Power and Propulsion Portfolio through a DURIP (FA9550-20-1-0191) and the Joint Advanced Propulsion Institute, a NASA Space Technology Research Institute.  
The authors would like to thank Dr. Zachariah Brown, Madison Allen, and William Hurley for experimental assistance, as well as Dr. Thomas Marks, Dr. Leanne Su, and Declan Brick for theoretical assistance.
\end{acknowledgements}

\bibliography{main.bib}

\end{document}